\pdfoutput=1
\documentclass[a4paper,12pt]{article}

\usepackage[utf8]{inputenc}
\usepackage[T1]{fontenc}
\usepackage[english]{babel}
\usepackage{amsmath}
\usepackage{mathtools}
\usepackage{comment}
\usepackage{textcomp}
\usepackage{siunitx}
\usepackage[style=base]{caption}
\usepackage{subfig}
\usepackage{bm}
\usepackage{url}
\usepackage{pdflscape}

\usepackage{setspace}
\usepackage[bitstream-charter]{mathdesign}
\usepackage{csquotes}
\usepackage[
    backend=biber,
    citestyle=numeric,
    bibstyle=phys,
    articletitle=false,
    biblabel=brackets,
    chaptertitle=false,
    pageranges=false,
    isbn=false,
    url=false,
    sorting=none
]{biblatex}
\DeclareFieldFormat{titlecase}{#1}
\bibliography{sources}

\newcommand{\mytitle}{Feedback-Control of Photoresponsive Fluid Interfaces}
\newcommand{\authors}{Josua Grawitter and Holger Stark}

\newcommand{\dd}{\mathrm{d}}
\newcommand{\dif}{\,\dd}

\newcommand{\eq}{\mathrm{eq}}
\newcommand{\funcdif}[2]{\frac{\delta #1}{\delta #2}}

\newcommand{\kbt}{k_\mathrm{B} T}
\newcommand{\rlim}{\rho_\mathrm{\scriptstyle \infty}}
\newcommand{\ofk}{(\vect{k})}
\newcommand{\ofr}{(\vect{r})}

\newcommand{\ofx}{(\vect{x})}
\newcommand{\partdif}[2]{\frac{\partial #1}{\partial #2}}
\newcommand\Mar{\mbox{\textit{Mg}}}
\newcommand{\vect}[1]{\bm{#1}}
\newcommand{\matr}[1]{\mathsf{#1}}

\usepackage[
    colorlinks=false,
    pdfborder={0 0 0},
    linktoc=all,
    pdfpagelabels,
    plainpages=false,
    bookmarks=true,
    hypertexnames=false, 
    pdftex,
    pdfauthor={\authors},
    pdftitle={\mytitle}
]{hyperref}
\usepackage{cleveref}
\usepackage{autonum}

\title{\mytitle}
\author{\authors}

\begin{document}
\renewcommand{\thefootnote}{\alph{footnote}}
\noindent
{\LARGE \textbf{Feedback-Control of Photoresponsive \newline Fluid Interfaces}}

\vspace{0.5cm}
\noindent \textit{
Josua Grawitter\footnotemark[1]\textsuperscript{,}\footnotemark[2]\textsuperscript{,}\footnotemark[3] and Holger Stark\footnotemark[1]\textsuperscript{,}\footnotemark[4]
}
\hfill October~25, 2017

\footnotetext[1]{Institut für Theoretische Physik, Technische Universität Berlin, Hardenbergstr.~36, 10623 Berlin, Germany.}
\footnotetext[2]{Department of Physics and Astronomy, University of Pennsylvania, 209~South 33rd Street, Philadelphia, PA 19104-6396, USA.}
\footnotetext[3]{E-mail:~\href{mailto:josua.grawitter@physik.tu-berlin.de}{\nolinkurl{josua.grawitter@physik.tu-berlin.de}}}
\footnotetext[4]{E-mail:~\href{mailto:holger.stark@tu-berlin.de}{\nolinkurl{holger.stark@tu-berlin.de}}}

\renewcommand{\thefootnote}{\arabic{footnote}}

\subsubsection*{Abstract}

Photoresponsive surfactants provide a unique microfluidic driving mechanism.
Since their molecular shapes change under illumination and thereby affect  surface tension of fluid interfaces, Marangoni flow along the interface occurs.
To describe the dynamics of the surfactant mixture at a planar interface, we formulate diffusion-advection-reaction equations for both surfactant densities.
They also include adsorption from and desorption into the neighboring fluids and photoisomerization by light.
We then study how the interface responds when illuminated by spots of light.
Switching on a single light spot, the density of the switched surfactant spreads in time and assumes an exponentially decaying profile in steady state.
Simultaneously, the induced radial Marangoni flow reverses its flow direction from  inward to outward.
We use this feature to set up specific feedback rules, which couple the advection velocities sensed at the light spots to their intensities.
As a result two neighboring spots switch on and off alternately.
Extending the feedback rule to light spots arranged on the vertices of regular polygons, we observe periodic switching patterns for even-sided polygons, where two sets of next-nearest neighbors alternate with each other.
A triangle and pentagon also show regular oscillations, while heptagon and nonagon exhibit irregular oscillations due to frustration.
While our findings are specific to the chosen set of parameters, they show how complex patterns at photoresponsive fluid interfaces emerge from simple feedback coupling.


\newpage

\section{Introduction}\label{intro}

New microfluidic devices require a better understanding of fluids and flow control on a microscopic scale.
For example, advances in digital and inertial microfluidics have resulted in useful biochemical reactors~\cite{samiei_review_2016} and precise sorting for micron-sized objects~\cite{zhang_fundamentals_2016}.
These devices often rely on specific geometries to control fluid flow~\cite{huang_continuous_2004,prohm_optimal_2013,otto_realtime_2015,schaaf_inertial_2017}.
But flow can also result from gradients in surface tension through a phenomenon known as the \emph{Marangoni effect}.
Gradients of surface tension at the interface of fluids result from curvature~\cite{sternling_interfacial_1959,mokbel_influence_2017} or temperature~\cite{garnier_optical_2003,gugliotti_surface_2004} gradients, or from gradients in the concentration~\cite{thomson_certain_1855,blawzdziewicz_stokes_1999,hanczyc_fatty_2007,thutupalli_swarming_2011} of \emph{surfactants}, i.e.~molecules which attach to fluid interfaces and reduce surface tension.
Surfactants containing one or more stereoisomers, like azobenzene~\cite{shinkai_photocontrol_1982}, spiropyran~\cite{sakai_photoisomerization_2007}, or stilbene~\cite{eastoe_properties_2002} are \emph{photoresponsive}:
They reversibly change shape under illumination with specific wavelengths of light and thereby interact differently with bordering fluids and neighboring surfactants~\cite{shang_photoresponsive_2003}.
Consequently, their different shapes affect surface tension differently and lead to Marangoni advection between illuminated and dark regions of the interface~\cite{chevallier_pumping_2011,schmitt_marangoni_2016}.
Through this mechanism, photoresponsive surfactants offer a unique way to control fluid flow by shining static or changing patterns of light on an interface~\cite{baigl_photo_2012,kavokine_light_2016}.

Various applications for photoresponsive surfactants have been reported:
They have been used to create responsive foams~\cite{chevallier_photofoams_2012,fameau_responsive_2015}, fluids with tunable viscosity~\cite{lee_photoreversible_2004}, and photoresponsive polymer brushes~\cite{kopyshev_making_2015}.
Beyond tunable materials, they allow to move and position individual surfactant-covered droplets~\cite{diguet_photomanipulation_2009}, to turn droplets into self-propelled particles~\cite{schmitt_marangoni_2016}, and to pump fluid near a planar water surface~\cite{chevallier_pumping_2011}.
The transport mechanisms in all three examples are based on the interplay of two or more processes determining the dynamics of photoresponsive surfactants at fluid interfaces.

\begin{figure}
\centering
\includegraphics[width=8.1cm]{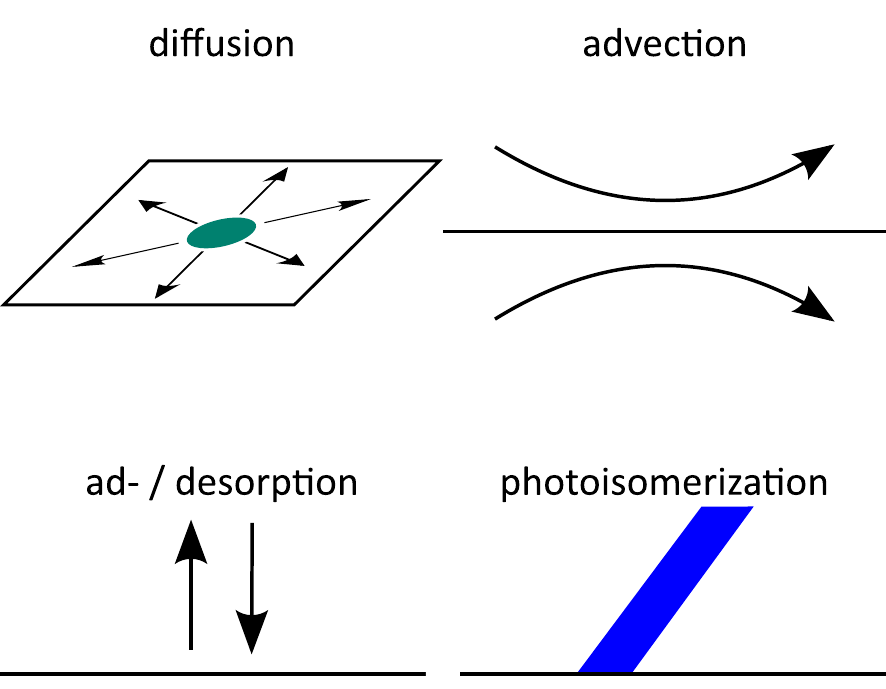}
\caption{Schematics of all four processes governing surfactant motion at a fluid-fluid interface.}
\label{interface}
\end{figure}

Here, we consider the four processes displayed schematically in Fig.~\ref{interface}:
Firstly, surfactants diffuse at the interface thermally and drift down chemical gradients.
Secondly, surfactants at the interface are advected by Marangoni flow set up by gradients in surface tension in the bordering fluids.
Thirdly, surfactants occasionally desorb from the interface into one of the fluids or adsorb to the interface from a fluid.
And fourthly, surfactants photoisomerize or switch molecular shape, when they are exposed to light of a specific wavelength.
All four processes are rationalized in two diffusion-advection-reaction equations for the surfactant densities.
Earlier theoretical descriptions~\cite{zinemanas_viscous_1987,nadim_concise_1996,pozrikidis_interfacial_2001,schmitt_swimming_2013} have not considered photoisomerization, or in the case of Ref.~\cite{schmitt_marangoni_2016} considered it only in the high-intensity limit.
However, because these processes happen concurrently, they couple to each other and produce complex behaviors:
For example, photoisomerization affects surface tension, which induces advection, and also leads to concentration gradients and diffusion.
By integrating all processes in our theory, we are able to investigate these complex couplings.

While previous research has focused on manipulating droplets~\cite{moeller_controlling_1998,diguet_photomanipulation_2009,schmitt_marangoni_2016}, we use the relative
simplicity of a planar interface to generate and study more complex flow and light patterns in theory.
In particular, we aim at demonstrating that the method of feedback control~\cite{astrom_feedback_2008}, where some inherent dynamic quantity is sensed and coupled back to the system, can be used to design fluid flow and induce novel dynamic patterns.

In this article, we specifically study the effect of point-like spots of light on a planar fluid-fluid interface with photoresponsive
surfactants for a specific set of realistic parameters.
Because these surfactants usually respond to specific wavelengths of light, one can easily imagine a laser beam or similarly focused light source generating these spots~\cite{baigl_photo_2012}.

When a single light spot is switched on, it generates a dynamic response where the density of the switched surfactant spreads in time and assumes an exponentially decaying profile in steady state.
Simultaneously, the induced radial Marangoni flow reverses its flow direction from inward to outward.
We use this feature to set up a specific feedback rule based on advection velocities sensed at the light spots.
They couple back to the on/off states of the light spots and thereby initiate self-generated oscillations of two neighboring spots.
We generalize our approach to regular polygons.
While we mostly observe regular oscillatory patterns, seven and nine-sided polygons exhibit irregular oscillations due to frustration.
The  simple on/off switching of the light spots is similar to thermostat control, which typically leads to oscillatory patterns~\cite{friedman_periodic_1988,gurevich_periodicity_2009} and has been applied to microfluidic problems before~\cite{prohm_feedback_2014,zeitz_feedback_2015}.

The article is organized as follows:
We describe our theory in section~\ref{sec_theory} and the numerical method to solve the coupled dynamic equations in section~\ref{sec_solver}.
In section~\ref{sec_single} we investigate the response to a single light spot switching on and off.
Section~\ref{sec_coupling} presents two light spots coupled by feedback control and section~\ref{sec_polygons} the extension to regular polygons with up to 10 edges.
We close with a summary and conclusions in section~\ref{sec_conclusions}.

\section{Theory}
\label{sec_theory}

There is a long tradition of describing surfactant covered interfaces with  diffusion-advection-reaction equations \autocite{zinemanas_viscous_1987,nadim_concise_1996,pozrikidis_interfacial_2001,schmitt_swimming_2013} governing the coupled dynamics of surfactant densities $\rho_q$,
\begin{equation}
\partdif{\rho_q}{t}
= -\nabla\cdot (\vect{j}_q^\mathrm{D} + \rho_q\vect{v}^\mathrm{A})
  + S_q^\mathrm{ads} + S_q^\mathrm{iso} \, .
\label{eq_ansatz}
\end{equation}
Here, the diffusive currents~$\vect{j}_q^D$ result from thermal motion and local interactions of the surfactants, the advection velocity~$\vect{v}^A$ from drift due to the Marangoni effect, and the reaction terms~$S_q^\mathrm{ads}$ describe ad- and desorption to and from a fluid-fluid interface, respectively.
In addition, we introduce reaction terms~$S_q^\mathrm{iso}$ to describe photoisomerization.
Our first objective is to close this system of partial differential equations by expressing all four r.h.s.~terms of Eq.~(\ref{eq_ansatz}) as functions of the respective densities of isomers A and B, which we denote $\rho_A$ and $\rho_B$.
In the following we describe all terms in more detail, formulate the equations in reduced units, and give relevant values for all material parameters.

\subsection{Free energy}

Motivated by the Flory-Huggins theory \cite{huggins_solutions_1941,flory_thermodynamics_1941}, we start from a coarse-grained Helmholtz free energy~$F$ of a compressible mixture of two components, $A$ and $B$, with mean-field interactions.
We define $F$ as the integral over a free energy density~$f$ on the interface, such that $F = \int f\ofx \dif^2 \vect{x}$.
The first term of $f$ contains entropic contributions of each component (see e.g.~Chapter~3.1 in Ref.~\cite{hansen_theory_2006}), the second term describes mean-field interactions between the surfactants of the same or different type, and the final term is the intrinsic energy density~$\sigma_0$ of a clean interface without surfactants:
\begin{multline}
f =
\kbt \left[ \rho_A (\ln(\lambda^2 \rho_A) - 1)
+ \rho_B (\ln (\lambda^2\rho_B) - 1) \right]\\
+ \frac{1}{\rlim} \left(\frac{1}{2} E_A \rho_A^2 + E_{AB} \rho_A \rho_B
+ \frac{1}{2} E_B \rho_B^2 \right) + \sigma_0 \, .
\label{eq_energy}
\end{multline}
Entropic terms are proportional to temperature $\kbt$ and contain the thermal de Broglie wavelength~$\lambda$, which does not affect dynamics in our case.
Interaction terms are characterized by three energy constants~$E_A$, $E_{AB}$ and $E_B$ and the saturation density~$\rlim$ of a fully covered interface.

In appendix~\ref{flory_huggins} we describe how Eq.~(\ref{eq_energy}) relates to Flory-Huggins theory.

\subsection{Diffusive currents}

To formulate diffusive currents, we apply dynamical density functional theory (DDFT) as originally derived by Marconi and Tarazona \autocite{marconi_dynamical_1999} for colloids and generalized to binary mixtures by Archer \autocite{archer_dynamical_2005}.
The original derivation builds on a variant of the $N$-particle Smulochowski equation for a mixture of particles of type $q$ and density $\rho_q$.

By following the standard procedure, Archer arrives at a set of continuity equations.
They are coupled via current densities~$\vect{j}_q$, which depend on functional derivatives of $F$:
\begin{equation}
 \vect{j}_q\ofx
 = -\frac{1}{\gamma_q} \rho_q\ofx
  \nabla \funcdif{F[\rho_A, \rho_B]}{\rho_q\ofx}
 = -\frac{1}{\gamma_q} \rho_q\ofx
  \nabla \mu_q\ofx \, .
  \label{eq_diffusion}
\end{equation}
We calculate them for each surfactant species and assume identical friction coefficients $\gamma_A=\gamma_B=\gamma$.
Using the free energy density $f$ from Eq.~(\ref{eq_energy}) and introducing the diffusion constant $D= \kbt / \gamma$, we arrive at
\begin{align}
 \vect{j}_A^\text{D}
 &= -D \left[\nabla \rho_A
 + \frac{\rho_A}{\kbt \rlim}
 \left( E_{A} \nabla \rho_A
  + E_{AB} \nabla \rho_B \right) \right]\\
 \vect{j}_B^\text{D}
 &= - D \left[ \nabla \rho_B
 + \frac{\rho_B}{\kbt \rlim}
 \left( E_{B} \nabla \rho_B
  + E_{AB} \nabla \rho_A \right) \right]
\end{align}
Linear terms represent conventional thermal diffusion and nonlinear terms result from interactions between surfactants.

\subsection{Advection: Marangoni effect}

Another important contribution to the current densities is advection due to Marangoni flow initiated by gradients in surface tension.
Here, we derive Marangoni currents for a planar fluid-fluid interface.

Surface tension~$\sigma$ is the thermodynamic conjugate variable to surface area, i.e.~$\sigma=\partial F / \partial A$.
In appendix~\ref{appendix_surface_tension} we show that for any free energy of the form $F=\int_A f(\rho_A,\rho_B) \dif^2\vect{x}$
\begin{equation}
\sigma = f - \rho_A \frac{\partial f}{\partial \rho_{A}} - \rho_B \frac{\partial f}{\partial \rho_{B}} \, .
\label{eq_tension_general}
\end{equation}
Like $f$, $\sigma$ is based on the usual adiabatic approximation of local equilibrium.
We substitute $f$ from Eq.~(\ref{eq_energy}) and arrive at
\begin{equation}
 \sigma = \sigma_0 -\kbt (\rho_A + \rho_B)
 - \frac{1}{\rlim} \left( \frac{1}{2} E_A \rho_A^2 + E_{AB} \rho_A \rho_B + \frac{1}{2} E_B \rho_B^2 \right) \, .
 \label{eq_surface_tension}
\end{equation}
Intuitively, the first two terms mean any surfactant lowers surface tension equally but the third term takes into account their different interaction energies.
For example, repulsive interactions ($E_i>0$) stabilize the interface by
lowering surface tension.

We assume small Reynolds numbers and therefore need to solve the Stokes equations for the flow field of an incompressible fluid characterized by pressure~$p$ and velocity $\vect{v}=(v_x, v_y, v_z)$:
\begin{equation}
 \nabla_{\vect{r}}[\eta(z)\nabla_{\vect{r}} \vect{v}] - \nabla_{\vect{r}} p = -\delta(z)\nabla_{\vect{x}}\sigma
\quad \text{and} \quad
\nabla_{\vect{r}}\cdot\vect{v}=0 \, ,
\label{stokes_eq}
\end{equation}
where the viscosities of the two bounding fluids are
\begin{equation}
\eta(z>0) = \eta_{+}
\quad\text{and}\quad
\eta(z<0) = \eta_{-} \, .
\end{equation}
The interface is at $z=0$ and the gradient of surface tension~$\sigma$ acts on both fluids like an external force.
The vector~$\vect{r}=(x,y,z)$ refers to any position in three dimensions, while $\vect{x}=(x,y,0)$ is restricted to positions on the interface.
Similarly, we define gradient operators $\nabla_{\vect{r}} = (\partial_x, \partial_y, \partial_z)$ and $\nabla_{\vect{x}}=(\partial_x, \partial_y, 0)$.
We do not allow flow through the interface and impose vanishing velocity and pressure far from the interface:
\begin{align}
v_z(z=0)&=0 \, , &
\vect{v}(|z|\to\infty)&=0 & \! \! \text{and} \,\;
p(|z|\to\infty)&=0
\nonumber
\end{align}

\begin{figure}
\centering
\includegraphics[width=4.5cm]{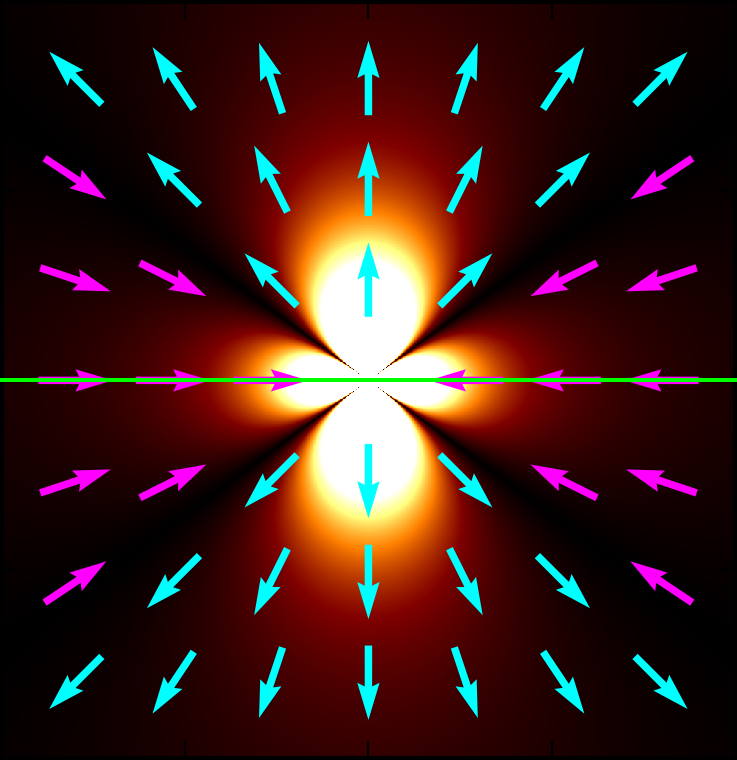}
\caption{Fluid velocity field initiated by a positive point-like inhomogeneity of the surface tension on a horizontal interface (green) separating two fluids.
Arrows indicate direction and background brightness indicates speed.
Close to the interface fluid is pulled towards the origin and pushed out along the normal.}
\label{response}
\end{figure}

The solution of Eq.~(\ref{stokes_eq}) can be computed using the modified Oseen tensor~$\matr{O}\ofr$ of Jones \textit{et al.}~to treat external forces acting along a fluid-fluid interface~\cite{jones_diffusion_1975,dominguez_collective_2016}:
\begin{equation}
\matr{O}(\vect{r}) = \frac{1}{8\pi\bar\eta |\vect{r}|} \left( \matr{I} + \frac{\vect{r}\otimes\vect{r}}{|\vect{r}|^2} \right) \, .
\end{equation}
The tensor depends only on the average viscosity $\bar\eta = (\eta_{+} + \eta_{-})/2$.
The velocity field~$\vect{v}$ initiated by an arbitrary surface tension profile~$\sigma$ is
\begin{equation}
\vect{v}(\vect{r}=\vect{x} + z\vect{e}_z)
= \int \matr{O}(\vect{x - x'} + z\vect{e}_z) \nabla_{\vect{x'}} \sigma(\vect{x'}) \dif^2\vect{x'} \, .
\label{eq_oseen}
\end{equation}
We apply the product rule, Gauss's theorem and show in appendix~\ref{app_oseen} that the resulting ring integral vanishes for
a constant value~$\sigma=\sigma_\eq$ at the boundary.\footnote{Note that $\sigma_\eq$ refers to an interface in equilibrium while $\sigma_0$ refers to a clean interface; their values are usually different.}
Thus, we arrive at the velocity field
\begin{equation}
\vect{v}\ofr
= \int
  \vect{g}(\vect{r - x'})
  \sigma(\vect{x'})
\dif^2\vect{x'} \, ,
\label{eq_vectorform}
\end{equation}
where the velocity response function $\vect{g}\ofr$ is a three-di\-men\-sional vector field, which decays like $|\vect{r}|^{-2}$:
\begin{equation}
\vect{g}\ofr
= -\nabla_{\vect{x}}\cdot\matr{O}(\vect{r})
= -\frac{1}{8\pi\bar\eta} \left(
  \frac{\vect{r}}{|\vect{r}|^3}
  - 3z^2 \frac{\vect{r}}{|\vect{r}|^5}
  \right) \, .
\end{equation}
Figure~\ref{response} displays $\vect{g}$ in a plane perpendicular to the interface and containing the origin.

For our numerical solutions of Eq.~(\ref{eq_ansatz}) we only need $\vect{v}$ at the interface, i.e.~at $z=0$.
There, $\vect{v}\ofx$ can be rewritten as a gradient field with $\delta\sigma\ofx=\sigma\ofx-\sigma_\eq$:
\begin{equation}
\vect{v}(\vect{x})
= \frac{1}{8\pi\bar\eta} \nabla_{\vect{x}} \int
  \frac{\delta\sigma(\vect{x'})}{|\vect{x-x'}|}
  \dif^2\vect{x'} \, .
\label{eq_mainresult}
\end{equation}
This form of $\vect{v}$ requires less computational effort than Eqs.~(\ref{eq_oseen}) and (\ref{eq_vectorform}) because it minimizes the number of scalar convolutions we need to perform.
Note that the integral only converges if $\delta\sigma\ofx$ decays like $|\vect{x}|^{-\varepsilon}$, $\varepsilon > 1$.

As a sidenote, the integral in Eq.~(\ref{eq_mainresult}) is a \textit{Riesz potential}~\autocite{riesz_integrale_1949}, a form of the inverse half Laplacian~$(-\Delta_{\vect{x}})^{-1/2}$ operating on $\delta\sigma$.
The operator is defined by its Fourier transform, such that
\begin{equation}
(-\Delta_{\vect{x}})^{-1/2} f\ofx
= \mathcal{F}^{-1}\left( \frac{\mathcal{F}(f)\ofk}{|\vect{k}|} \right)
= -\frac{1}{2\pi} \int \frac{ f (\vect{x'})}{|\vect{x-x'}|}  \dif^2\vect{x'}
\end{equation}
where $\mathcal{F}$ denotes the two-dimensional Fourier transform, and $\mathcal{F}^{-1}$ its inverse.

\subsection{Ad- and desorption}

We use the Frumkin approach to adsorption, which describes adsorbed particles as a real gas and desorbed particles as an
ideal gas~\autocite{frumkin_beeinflussung_1926}.
Because it takes into account surface coverage and local interactions on the interface, it is an extension of the Langmuir adsorption model \autocite{langmuir_adsorption_1918}.
Adsorption processes add two nonconservative terms to our differential equations (\ref{eq_ansatz}): one for adsorption, which is proportional to local free space on the interface, and one for desorption, which is proportional to the local density of each component and contains an Arrhenius factor to account for the changing surface tension or energy.
Thus, the reaction terms in Eqs.~(\ref{eq_ansatz}) read
\begin{equation}
S_q^\mathrm{ads}(\rho_A, \rho_B)
= k^\text{a}_q \cdot (\rho^\infty - \rho_A - \rho_B)
- k^\text{d}_q \exp\left( -\frac{1}{\kbt}\partdif{\sigma}{\rho_q} \right) \rho_q \, ,
\label{eq_adsorption}
\end{equation}
where we also introduced adsorption and desorption rates $k^\text{a}_q$ and $k^\text{d}_q$ for each species.
The Arrhenius energy barrier for removing particles is given by the corresponding change in surface tension,
i.e.~$\partial_{\rho_q} \sigma$~\autocite{ferri_soluble_1999}.
The adsorption rate $k_q^\text{a}$ depends on the concentration of surfactants in the fluids and is assumed to be tunable in experiment.
In thermal equilibrium the reaction term vanishes and one obtains the equilibrium densities~$\rho^\eq_q$ by solving
\begin{equation}
\begin{pmatrix}
S_A^\mathrm{ads}(\rho^\eq_A,\rho^\eq_B)\\[1ex]
S_B^\mathrm{ads}(\rho^\eq_A,\rho^\eq_B)
\end{pmatrix}
= 0
\label{eq_ads_eq}
\end{equation}
for given rate ratios $k_A^\text{a}/k_A^\text{d}$ and $k_B^\text{a}/k_B^\text{d}$.

\subsection{Photoisomerization}

We describe photoisomerization using rate equations with local isomerization rates $k_{i \to j}$ in each direction:
\begin{align}
S_A^\mathrm{iso}(\rho_A, \rho_B)
&= k_{\scriptstyle B \to A} \rho_B
    - k_{\scriptstyle A \to B} \rho_A\\
S_B^\mathrm{iso}(\rho_A, \rho_B)
&= k_{\scriptstyle A \to B} \rho_A
    - k_{\scriptstyle B \to A} \rho_B
\label{eq_isomerization}
\end{align}
For our specific examples in Sec.~\ref{sec_results} we only allow for photoinduced isomerization from $A$ to $B$ surfactants and therefore set $k_{\scriptstyle B \to A} = 0$.
Note that total surfactant density is conserved during photoisomerization, which is quantified by $S_A^\mathrm{iso}+S_B^\mathrm{iso}=0$.

The isomerization rates in our examples are spatially inhomogeneous.
For example, a set of $N$~point-like light spots at positions~$\vect{x}_l$ is described by a superposition of Dirac-$\delta$~func\-tions with isomerization functions~$K_l(t)$:
\begin{equation}
k_{i \to j}(\vect{x}, t) = \sum\limits_{l=1}^N K_l (t) \delta(\vect{x}-\vect{x}_l) \, .
\label{eq_iso_spots}
\end{equation}
Each $K_l$ is proportional to the intensity of its light spot~$l$.
Thus, when the light spot is turned on, $K_l$ switches from 0 to a constant value~$K_{A \to B}$ specified in Table~\ref{table_parameters}.

\subsection{Closed and reduced equations}
\label{sec_reduced}

We nondimensionalize all quantities using $\kbt$ for energies, desorption rate $k^\text{d}_A$ for inverse time, saturation density $\rlim$ for densities, and $L = \sqrt{D / k^\text{d}_A}$ for length scales.
Note that $L$ is the length an $A$ surfactant diffuses until it desorbs from the interface.
In appendix~\ref{app_char_quant} we explain further the origin of these characteristic quantities, in particular, the rescaling of length and time.

The Marangoni number quantifies the importance of Marangoni flow by comparing the diffusion time for the characteristic length $L$, $t_{\mathrm{diff}} = L^2 /D$, to the time for advection by Marangoni flow,
$t_{\mathrm{Mg}} = L / v_{\mathrm{Mg}} = L \Delta \sigma / \bar{\eta}$.
Thus,
\begin{equation}
\Mar
= \frac{t_{\mathrm{diff}}}{t_{\mathrm{Mg}}}
= \frac{L \Delta \sigma}{\bar{\eta}D}
= \frac{\kbt \rlim}{\bar\eta \sqrt{k^\text{d}_A D}}
\label{eq_marangoni}
\end{equation}
To arrive at the last expression, we set $\Delta \sigma = \kbt \rlim$ and used the definition for $L$.

Non-dimensionalizing Eqs.~(\ref{eq_ansatz}), (\ref{eq_diffusion}), (\ref{eq_mainresult}), (\ref{eq_adsorption}), and (\ref{eq_isomerization}) leads to reduced equations for each component~$q$:
\begin{multline}
    \partdif{\rho_q}{t} = \nabla \cdot \left(
    \rho_q \nabla \mu_q
    + \frac{1}{4} \Mar \rho_q \nabla (-\Delta)^{-\frac{1}{2}} \delta\sigma \right)\\
    + k^\text{a}_q \cdot (1 - \rho_A - \rho_B)
    - k^\text{d}_q \rho_q \exp\left(-\partdif{\sigma}{\rho_q}\right)
    \mp \; k_{\scriptstyle A \to B} \rho_A
\label{eq_nondim}
\end{multline}
with
\begin{align}
\mu_A( \rho_A, \rho_B)
 &= \funcdif{F}{\rho_A}
 = \ln  \rho_A  + E_A \rho_A + E_{AB} \rho_B + \text{const.}
\nonumber\\
\mu_B( \rho_A, \rho_B)
 &= \funcdif{F}{\rho_B}
 = \ln  \rho_B  + E_B \rho_B + E_{AB} \rho_A + \text{const.}
\nonumber\\
\sigma( \rho_A, \rho_B)
 &= \sigma_0 - (\rho_A +  \rho_B)
    - \left( E_{AB}  \rho_A  \rho_B + \frac{1}{2}  E_A  \rho_A^2
    + \frac{1}{2}  E_B  \rho_B^2 \right)
\end{align}
and the signs $-$ and $+$ of the photoisomerization term refer to the $A$ and $B$ surfactants, respectively.

Our numerical solutions are calculated on a square region~$\mathcal{B}$ of the interface, where we solve Eq.~(\ref{eq_nondim}) with open, Dirichlet-like boundary conditions:
We set the densities~$\rho_q$ to their equilibrium values~$\rho^\eq_q$ at the boundary but also outside $\mathcal{B}$, so that
\begin{equation}
\delta\sigma(\vect{x} \notin \mathcal{B})
= \sigma(\rho_A^\eq, \rho_B^\eq) - \sigma_\eq
= 0 \, .
\end{equation}
Thereby, the region of integration of the integral in Eq.~(\ref{eq_mainresult}) is restricted to~$\mathcal{B}$.

\subsection{Parameters}

The reduced equations contain parameters which are determined by material properties.
Our estimates for these parameters are collected in Table~\ref{table_parameters}.
\begin{table}[tb]
\centering
\begin{tabular*}{0.4\textwidth}{@{\extracolsep{\fill}}cl}
\hline
parameter & value \\
\hline
$E_A$ & $1.5\,\kbt$ \\
$E_{AB}$ & $1.3\,\kbt$ \\
$E_B$ & $1.1\,\kbt$ \\
$\rho^\eq_A$ & $0.5 \,\rlim$ \\
$\rho^\eq_B$ & $0.0 \,\rlim$ \\
$\Mar$ & $1.2 \cdot 10^5$ \\
$k^\text{d}_B$ & $1\,k^\text{d}_A$ \\
$k^\text{a}_A$ & $2.12\,k^\text{d}_A$ \\
$k^\text{a}_B$ & $1.73\,k^\text{d}_A$ \\
$K_{A \to B}$ & $10\,D$ \\
\hline
\end{tabular*}
\caption{Our estimates for various material and photo\-isomerization parameters.
The values for $k^\text{a}_{A/B}$ are calculated using Eq.~(\ref{eq_ads_eq}).}
\label{table_parameters}
\end{table}
These estimates are not meant to replicate a specific combination of fluids and surfactants but rather to model a realistic system, which exhibits stable mixtures of surfactant species on the fluid interface.

The values for the interaction energies $E_A$, $E_B$, and $E_{AB}$ are based on data for surface tensions of surfactant mixtures collected by Chevallier \textit{et al.}~for AzoTAB at a water-air interface \autocite{chevallier_pumping_2011} assuming $E_{AB}=(E_A + E_B)/2$ (for details see appendix \ref{parameter_extraction}).
Our choice of $\Mar = 1.2\cdot10^5$ is also based on their data, assuming $D=\SI{900}{\micro\meter^2 \second^{-1}}$, $\rlim=\SI{5}{nm^{-2}}$, $k^\text{d}_A=\SI{6.62}{s^{-1}}$, and $T=\SI{300}{K}$, and a larger value of the average viscosity $\bar{\eta}=\SI{2}{\milli\pascal\second}$.

The equilibrium surface densities $\rho^\eq_A$ and $\rho^\eq_B$ are assumed to be fully adjustable by changing the partial pressure of surfactants in the fluids, and for simplicity, we set $\rho^\eq_B=0$ and $\rho^\eq_A = 0.5\,\rlim$.
For the same reason we assume identical desorption rates $k^\mathrm{d}_B = k^\mathrm{d}_A$ for both isomers.
\footnote{Note that this is not valid for AzoTAB at an air-water interface as demonstrated by Chevallier \textit{et al.}~\autocite{chevallier_pumping_2011}.}
For reference, the aforementioned material parameters give a characteristic adsorption-diffusion length $L \approx \SI{12}{\micro\meter}$.

Photoisomerization is assumed to be fully tunable by adjusting laser intensity and focus.
Throughout our study we set all photoisomerization rates using the value for $K_{A \to B}$ in Table~\ref{table_parameters} unless explicitely stated.

\section{Numerical solver}
\label{sec_solver}

We solve Eq.~(\ref{eq_nondim}) numerically using a combination of established algorithms implemented in the Julia programming language~\autocite{bezanson_julia_2017}.
There are three major steps, which can be treated separately from each other: integration in time, discretization in space and interpolation in space.
We discuss all three in the following paragraphs.
Readers not interested in numerical details can skip this section.

We use a finite volume discretization on a uniform grid.
Accordingly, we tile the interface using roughly $56.000$ to $127.000$ hexagons (depending on the problem) and approximate all functions $f$ by their values $f_i$ at the center $\vect{x}_i$ of each cell $i$
(similar to \autocite{schmitt_active_2016}).
If we know $f$ exactly, we approximate it by the spatial average over each cell $\mathcal{C}_i$ with area~$A$.
For example, the photoisomerization field~$k\ofx=K\delta(\vect{x - \tilde x})$ for a single light spot is discretized by
\begin{equation}
k_i
= \frac{1}{A} \int\limits_{\vect{x}\in\mathcal{C}_i}
    k\ofx
  \dif^2\vect{x}
= \begin{cases}
  \frac{K}{A} & \text{if } \vect{\tilde x} \in \mathcal{C}_i
  \\
  0 & \text{if not}
\end{cases}
\end{equation}

The main advantage of finite volume methods is that they reliably conserve particle number in discetized continuum equations.
To discretize the diffusion term, we use the conventional central difference method, which is $2$nd order accurate (see e.g.~Ref.~\autocite{pang_introduction_2010}).
To treat the advective term correctly, we need an interpolation scheme to determine density values at the boundaries between cells.
Here, we use an upwind-dif\-fe\-rence method developed by B.~Koren (see Chapter 5 in Ref.~\autocite{vreugdenhil_numerical_1993}) which is $2$nd to $3$rd order accurate (depending on smoothness).\footnote{More precisely, Koren's method is total variation diminishing in Harten's sense~\autocite{harten_high_1983} and robust in Sweby's sense~\autocite{sweby_high_1984}.}
It is characterized by its flux-limiting function~$\phi$:
\begin{equation}
\phi(r) = \max\left\{0,
    \min\left[2r,
      \min\left(\frac{1}{3} + \frac{2}{3}r, 2\right)
    \right]
  \right\}\, ,
\end{equation}
where $r$ is the upwind gradient ratio between cells $i$ and $i+1$
of the advected field~$\rho_q$:
\begin{equation}
r(\vect{x}_{i+1/2})
= \frac{\rho_q(\vect{x}_{i+1}) - \rho_q(\vect{x}_{i})}{\rho_q(\vect{x}_i) - \rho_q(\vect{x}_{x-1})} \, .
\end{equation}
The index $i+1/2$ indicates the boundary between cells $i$ and $i+1$.
For a complete description of the method we refer to Ref.~\autocite{vreugdenhil_numerical_1993}.

Equation~(\ref{eq_mainresult}) contains a nonlocal term which is a convolution of surface tension and $|\vect{x}|^{-1}$.
Rather than performing the direct convolution, which would need an $\mathcal{O}(N^2)$-operation on a mesh with $N$ cells, we compute the integral in Fourier space.
The fast Fourier transform (FFT) is an $\mathcal{O}(N \log N)$-operation on orthogonal meshes.
We use a regridding scheme to switch between hexagonal and rectangular cells and thereby take advantage of FFT (see appendix~\ref{app_regridding}).

We integrate in time using the adaptive Runge-Kutta-Fehl\-berg (RKF45) method~\autocite{fehlberg_klassische_1970}.
RKF45 calculates $4$th- and $5$th-order approximations of trajectories, estimates the integration error from their distance, and adapts its time step to keep it below a threshold.
Integration proceeds using the $5$th-order approximation.

In our case, a trajectory is characterized at every point by the vector $\bm{\rho}$, which contains both densities, $\bm{\rho}=(\rho_A, \rho_B)$, and exists at every point on the interface.
The RKF45 error function should be a metric on the $\rho_q$ function space.
We base our metric on the $L^1$-norm because for any physically valid $\rho_q$ it simply gives the particle number~$N_q$, which is dimensionless and necessarily exists.
The $L^1$ norm also has a straight-forward counterpart for discretized functions, the Manhattan vector norm.
Because each component $\rho_q$ should fulfill the threshold criterion of the RKF45 method independently, we take the maximum of their norms:
\begin{equation}
||\bm{\rho}\ofx||
= \max\limits_{q\in\{A,B\}} \left\{ ||\rho_q\ofx||_{1} \right\} \, .
\end{equation}
The integration error~$e$ is simply the norm of the point-wise difference between two states~$\bm{\rho}$ and $\tilde{\bm{\rho}}$:
\begin{equation}
e[\bm{\rho}, \tilde{\bm{\rho}}]
= ||\bm{\rho}\ofx - \tilde{\bm{\rho}}\ofx|| \, ,
\end{equation}
where $\bm{\rho}\ofx$, $\tilde{\bm{\rho}}\ofx$ refer to the  $5$th- and $4$th-order approximations, respectively.
All data were produced using a tolance of $e\leq 10^{-6}$.

\section{Results}
\label{sec_results}

\subsection{Single light spot}
\label{sec_single}

We start by perturbing the interface with a single point-like spot of light.
Specifically, we study how the interface responds to a spot switching on and off as well as how it approaches steady state.
Our observations will form the basis for coupling the dynamics of multiple light spots by feedback control in section~\ref{sec_coupling}.

\begin{figure}
\centering
\includegraphics[width=9cm]{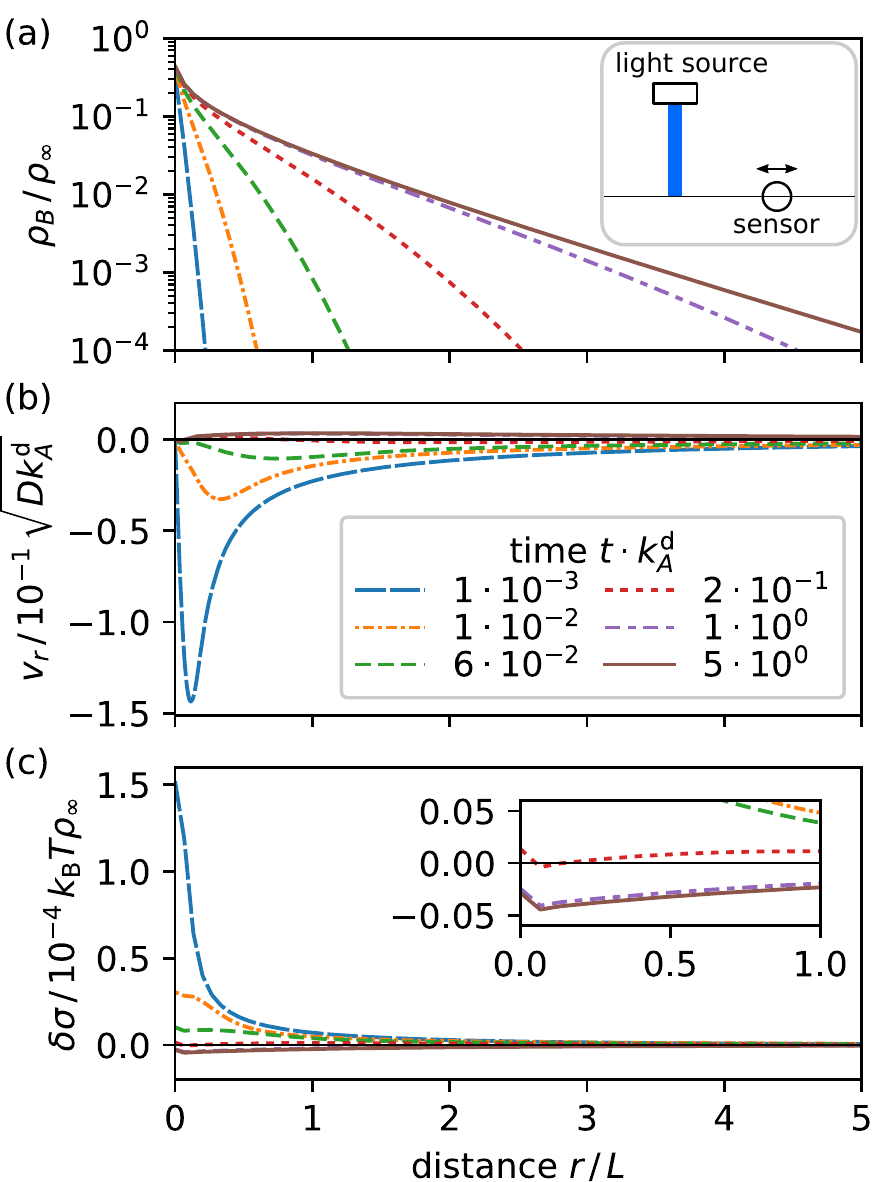}
\caption{
Radial profiles at various times~$t$ after a light spot at $r=0$ is switched on: (a)~surfactant density~$\rho_B$, (b)~radial velocity~$v_r$, and (c)~surface tension~$\delta\sigma$ relative to their equilibrium values.
Inset:~Blowup to show $\delta \sigma$ at the three latest times.
The solid line in each diagram shows the steady state nonequilibrium profile.
Emblem in (a):~The sensor senses $v_r$ and $\rho_B$ at a distance~$r$ from the light spot.
}
\label{single_transient}
\end{figure}

We solve Eqs.~(\ref{eq_nondim}) on a square region of area $20\,L \times 20\,L$ with a single light spot placed at its center.
Initially, only A surfactants are present.
They are homogeneously distributed and in adsorption equilibrium with the bordering fluids.
Then, the light spot is instantaneously turned on and switches surfactants from $A$ to $B$.
The resulting  radial profiles for density~$\rho_B$, radial velocity~$v_r$,  and surface tension~$\delta\sigma$ relative to their equilibrium values are displayed in Fig.~\ref{single_transient} for various times~$t$.

\begin{figure}[tp]
\centering
\includegraphics[width=9cm]{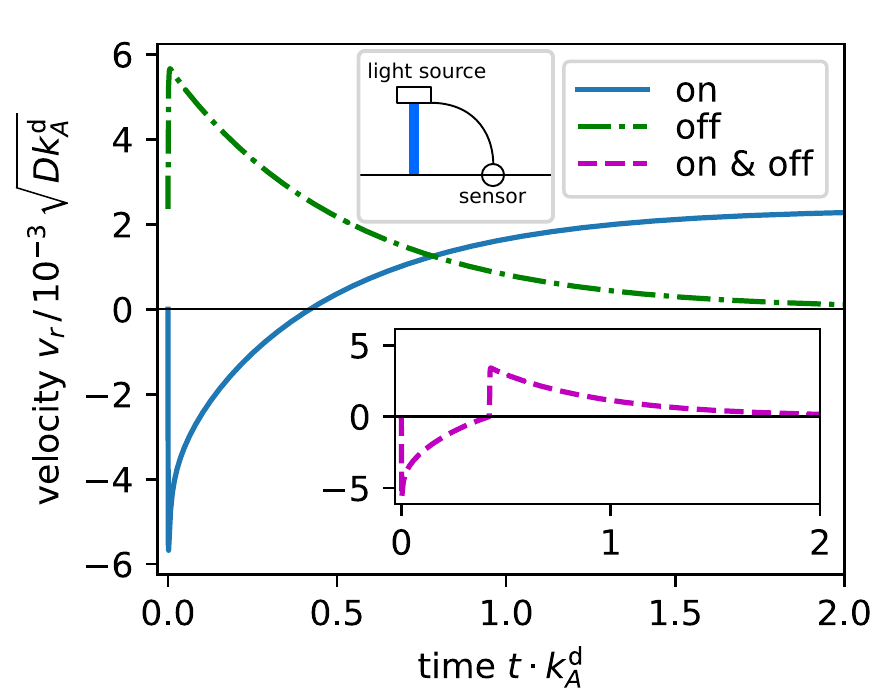}
\caption{
Radial velocity~$v_r$ at distance~$r=5\,L$ from the light spot plotted versus time~$t$ for light switched on (solid, blue) and switched off in the non-equilibrium steady state (dash-dotted, green).
Emblem:~A sensor measures $v_r$ initiated by the light spot and feeds it back to the light source.
Inset:~Radial velocity versus time, when the light spot is switched on at $t=0$ and switched off when $v_r$ crosses the zero line (dashed, purple).
}
\label{single_velocity_transient}
\end{figure}

As surfactants switch from $A$ to $B$, type $B$ surfactants start to diffuse radially away from the light spot and desorb~(a).
Simultaneously, a gradient in surface tension forms~(c) and starts to drive a Marangoni current~(b): at first toward the spot and later away from it.
This is illustrated in Fig.~\ref{single_velocity_transient}, where we plot the radial velocity at distance $r=5\,L$ from the spot: it continuously changes from pointing toward the spot (negative) to pointing away from it (positive) at $t\approx 0.5\,k^\mathrm{d}_A$.

Eventually, the surfactant densities approach a \emph{nonequilibrium} steady state in which adsorption, diffusion, advection, and photoisomerization balance each other.
In steady state, currents at the interface are nonzero because new type $B$ surfactants continue to diffuse and drift away from the light spot until they desorb and new type $A$ surfactants continue to replace switched surfactants by adsorption at the spot.
The radial density profile~$\rho_B(r)$ in Fig.~\ref{single_transient}(a) decays like $\mathrm{e}^{-\kappa r}$ with $\kappa \propto L$ as $r\to\infty$.\footnote{This is similar to the steady state of a simple diffusion equation with a point-source and constant global decay rate in two dimensions.
The analytical solution for that system is the $0$th order modified Bessel function of the second kind~$\mathcal{K}_0(r)$.
For large $r$ it scales as $r^{-1/2}e^{-r}$ (Eq.~(10.40.2) in Ref.~\cite{olver_nist_2017}).}

An interesting aspect of the current system is the reversal of the direction of the radial velocity.
Initially, Marangoni flow towards the light spot ($v_r < 0$) acts against the diffusive current of B surfactants, which in active emulsions can cause demixing (see Ref.~\autocite{schmitt_swimming_2013}).
However, here the Marangoni current ultimately aligns \emph{with} diffusion of type~$B$ surfactants ($v_r > 0$), while reaching steady state.
The directional change can be traced to an excess in the total surfactant density $\rho=\rho_A+\rho_B > \rho^\eq$, which develops with time in an increasing region around the light spot (not shown).
Because $\delta\sigma\approx-\kbt(\rho-\rho^\eq)$ in leading order, the surface tension becomes negative [see Fig.~\ref{single_transient}(c)] and the resulting positive slope determines the outward direction of $\vect{v}$.

In Fig.~\ref{single_velocity_transient} we display the time evolution of the radial velocity~$v_r$ at a specific distance~$r$, i.e.~the response of the system to switching the light spot on.
After reaching steady state, light is switched off and $v_r$ relaxes towards zero.
We now introduce a feedback control step in our system: when $v_r$ changes sign, we switch off the light spot.
The resulting time evolution of $v_r$ is plotted in the inset.
It gives an idea how we implement feedback control in section~\ref{sec_coupling}.

\begin{figure}[tp]
\centering
\includegraphics[width=9cm]{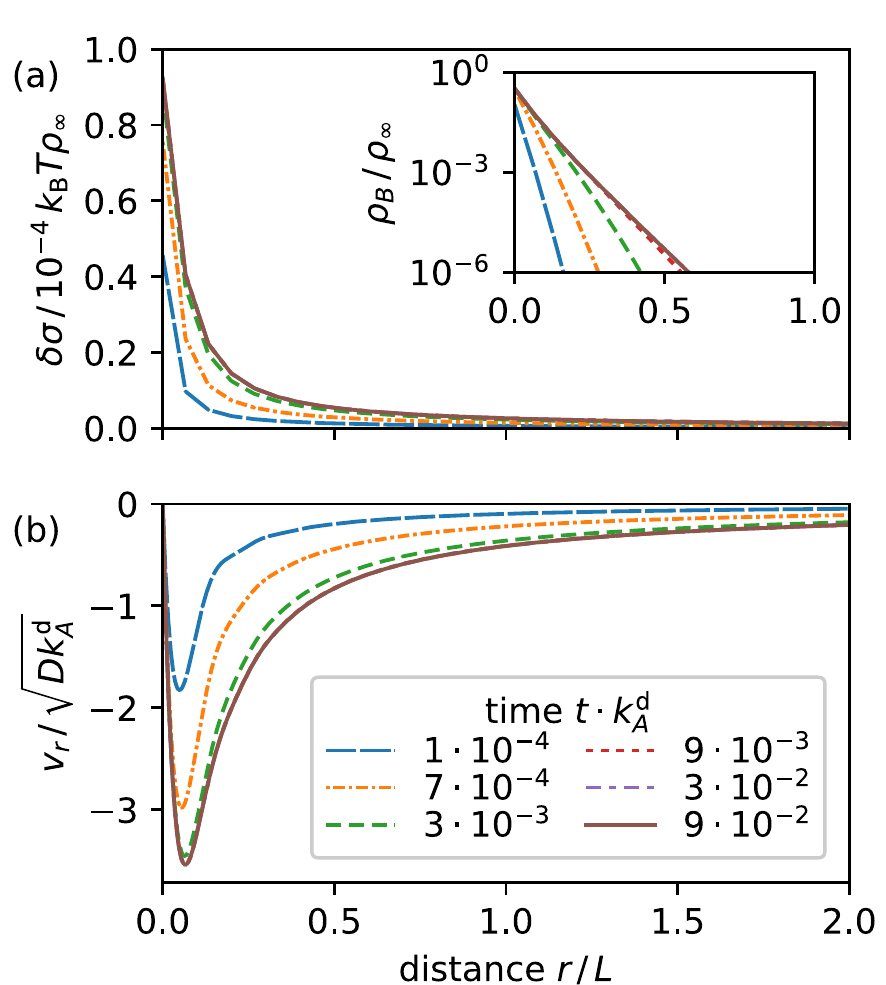}
\caption{
Radial profiles at various times~$t$ after a light spot at $r=0$ is switched on for $\Mar=6\cdot10^5$ and $k^\mathrm{d}_B=300\,k^\mathrm{d}_A$: (a)~surface tension~$\delta\sigma$ relative to equilibrium value (inset: surfactant density~$\rho_B$), and (b)~radial velocity~$v_r$.
The solid line in each diagram shows the steady-state nonequilibrium profile.
}
\label{single_transient2}
\end{figure}

For comparison, we include radial profiles in Fig.~\ref{single_transient2} for the case that the B surfactants desorb much faster than the A type ($k^\mathrm{d}_B=300\,k^\mathrm{d}_A$) as observed in Ref.~\cite{chevallier_pumping_2011}.
Now the surface tension is always positive.
It has a negative slope and the Marangoni current keeps its initial direction towards the light spot in steady state.
We attribute this simpler behavior to the increased desorption rate of type $B$, which prevents these surfactants from spreading across the interface.
As a result, diffusion becomes irrelevant and steady state is dominated by Marangoni currents driving type $A$ surfactants to the spot to replenish switched and desorbed surfactants there.
In the following, we concentrate on the more interesting case $k^\mathrm{d}_B = k^\mathrm{d}_A$.

\subsection{Two light spots coupled by feedback control}
\label{sec_coupling}

\begin{figure}
\centering
\includegraphics[width=9cm]{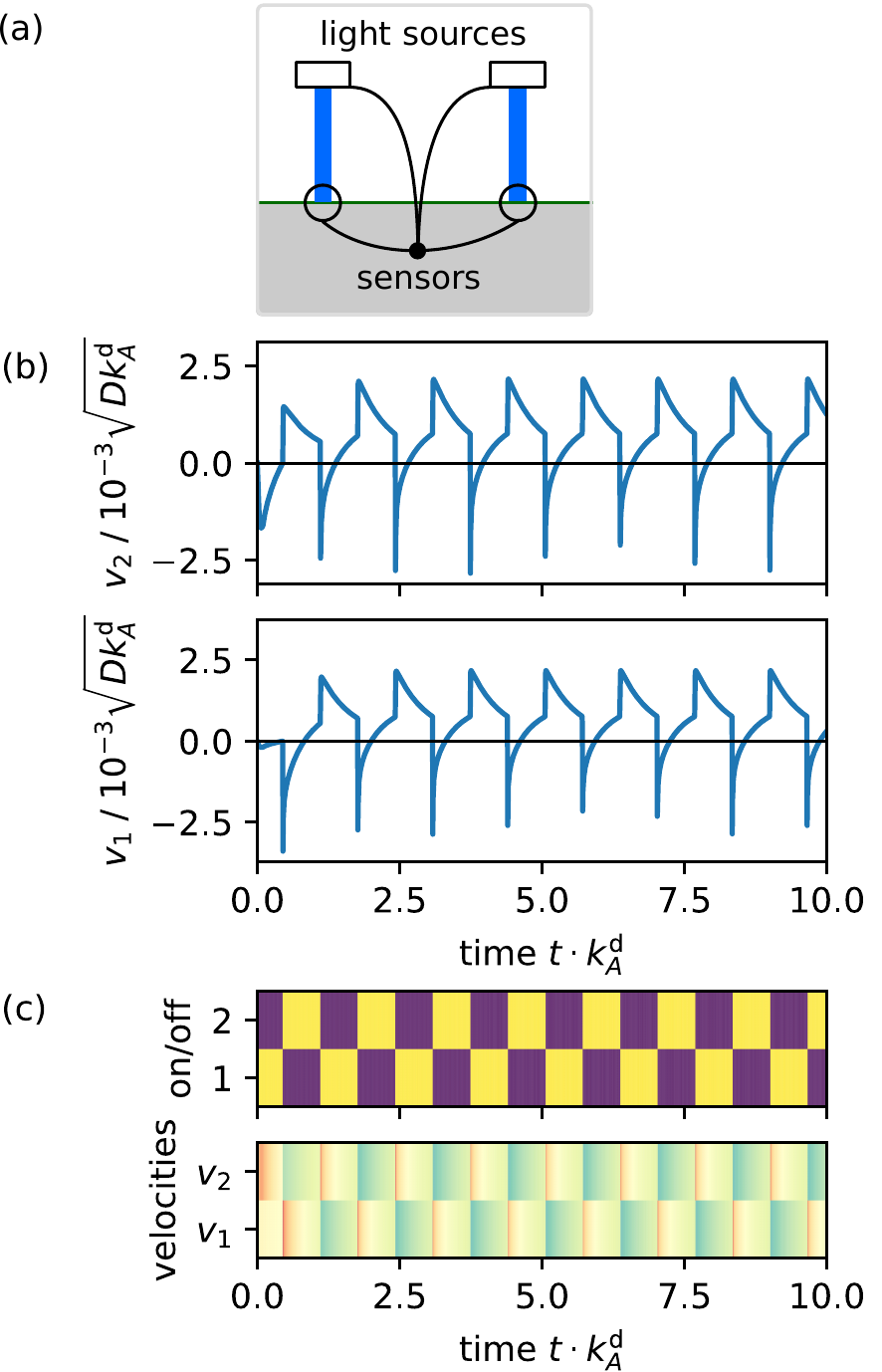}
\caption{
Example with two coupled light spots at a distance~$10\,L$ from each other:
(a)~Schematic of two light spots with their connected velocity sensors.
(b)~Velocities~$v_1$ and $v_2$ sensed at the locations of the respective light spots $1$ and $2$ plotted over time~$t$.
(c) Dynamic fingerprints of two light spots:
Color-coded time series for light spots (on: yellow/light, off: blue/dark) and velocities~$v_1$ and $v_2$ (advection towards
other spot: red, away from it: blue, darkness of shading indicates speed).
}
\label{two_laser_control2}
\end{figure}

The main idea is now that one senses the flow velocities at each light spot and specifies a feedback rule that couples back to the light spots and determines for each of them if they are switched on or off [see emblem in Fig.~\ref{two_laser_control2}(a)].
This means the system organizes its dynamic response by itself.

For two light spots a simple rule is that always the spot at which the velocity is largest is on and the other spot is off.
To rationalize this rule, we consider two light spots at respective positions~$\vect{x}_1$ and $\vect{x}_2$ with distance $\Delta x = |\vect{x}_1 - \vect{x}_2|$ [see schematic in Fig.~\ref{two_laser_control2}(a)].
In this setup the advection velocity $\vect{v}\ofx$ at one spot is always directed towards or away from the other light spot due to symmetry.
Accordingly, we define the advection velocity $v_l$ at one spot~$l$ to be positive, when it points away from the other spot and to be negative in the opposite case:
\begin{equation}
v_l
= \vect{v}(\vect{x}_l)
\cdot \frac{\vect{x}_l-\vect{x}_j}{\Delta x},
\quad j \neq l
\end{equation}
Only the spot with the larger~$v_l$ is switched on or equivalently, the isomerization function~$K_l(t)$ of each spot, which we introduced in Eq.~(\ref{eq_iso_spots}), is
\begin{equation}
K_l(t) = K_{A \to B} \theta \big(v_l(t) - v_j(t) \big), \quad j \neq l \, .
\end{equation}
Here, $\theta$ is the Heaviside step function and $K_{A \to B}>0$ is the isomerization constant while a light spot is switched on.
In the following, we always prepare the system in its equilibrium state and start by switching light spot~$1$ on at time~$t=0$.

Movie M01 in the Supplemental Material shows an animation how the density of the type~$B$ surfactant and the Marangoni flow velocities vary in time for two light spots with distance~$\Delta x=10\,L$.
In Fig.~\ref{two_laser_control2}(b) we display the velocities at the two spots, which govern the feedback dynamics.
As soon as the velocity~$v_2$ created by spot~$1$ at the location of spot~$2$ becomes positive, spot~$1$ is switched off and the active spot~$2$ generates a negative~$v_1$.
Over time the coupled light spots quickly establish regular oscillations in $v_1$ and $v_2$, which occur in antiphase.
In Fig.~\ref{two_laser_control2}(c) we represent the on-off state of the light spots and the velocities~$v_1$, $v_2$ in dynamic fingerprints, which we will use further to show the dynamic response of the system.

\begin{figure}
\centering
\includegraphics[width=9cm]{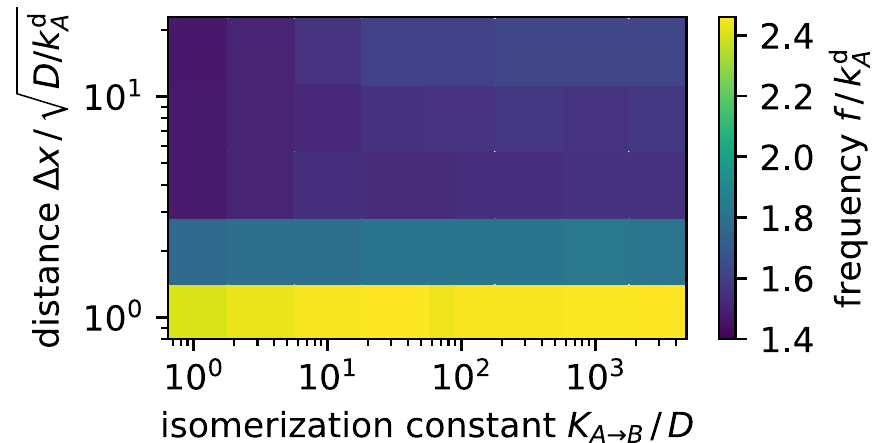}
\caption{Color-coded switching frequency~$f$ of two coupled light spots plotted versus distance~$\varDelta x$ and isomerization constant~$K_{A\to B}$.}
\label{two_laser_control}
\end{figure}

We observed the same switching behaviour of the light spots for a large set of distances and isomerization constants (proportional to light intensity).
The switching frequency~$f$ plotted in Fig.~\ref{two_laser_control} hardly depends on $K_{A \to B}$ in the studied parameter range of more than three decades.
Furthermore, increasing distance~$\Delta x$ by more than one decade reduces
$f$ by less than a factor of two.
Overall, $f$ is of the order of $k^\mathrm{d}_A$, which suggests the spots are coupled through a diffusion-desorption process.
An animation of the two alternating light spots is included as movie M02 in the Supplemental Material.

\subsection{Coupled light spots on regular polygons}
\label{sec_polygons}

\begin{figure}
\centering
\includegraphics[width = 4.5cm]{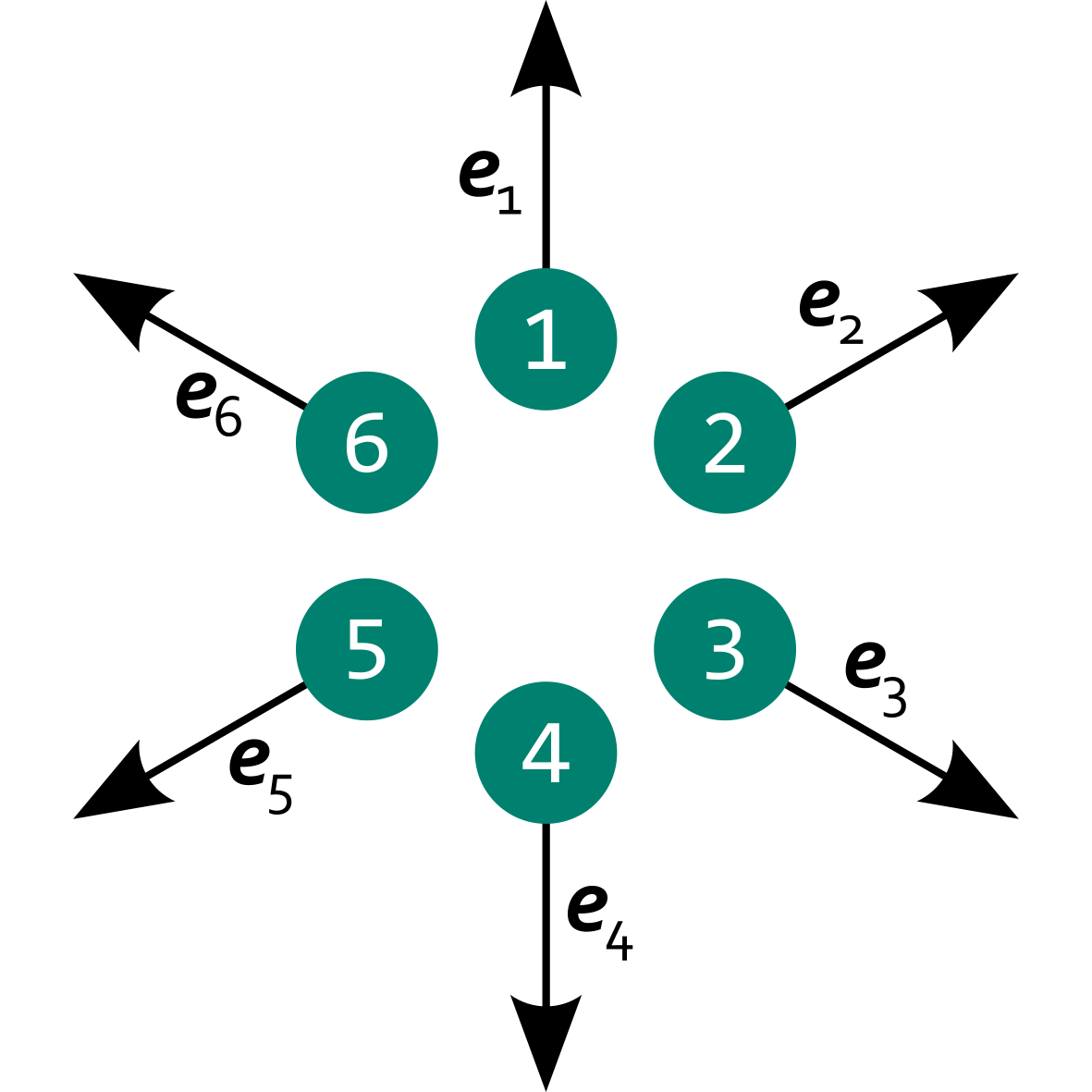}
\caption{Schematic of $6$ light spots positioned on the vertices of a regular hexagon, each with a radial projection vector $\vect{e}_l$ pointing away from the polygon's center.}
\label{schematic_polygon}
\end{figure}

Now, we generalize the previous situation:
Instead of two light spots we study $N$ spots arranged on the vertices of a regular polygon.
All polygons have the same circumscribed circle with diameter~$10\,L$ and the density and flow fields are computed on a square grid with a side length of $30\,L$.
The feedback-control rule is now formulated as follows.
Because the velocity at each spot does not have a predefined direction as in the case of $2$ spots, we project it onto the outward radial direction of the circumscribed circle as shown in Fig.~\ref{schematic_polygon} and arrive at a scalar velocity~$v_l$ for each spot~$l$:
\begin{equation}
v_l = \vect{v}(\vect{x}_l) \cdot \vect{e}_l \, .
\label{eq_projection}
\end{equation}
Each individual spot~$l$ is only kept switched on while $v_l$ is larger than the average of all $v_l$.
Equivalently, we introduce the isomerization function for spot~$l$:
\begin{equation}
K_l(t) = K_{A \to B} \theta\big(v_l (t) - \bar{v} (t) \big)
\end{equation}
with mean velocity $\bar{v} (t) = \frac{1}{N}\sum_{l=1}^N v_l (t)$ and the Heaviside step function~$\theta$.
If $N=2$ the condition simplifies to our feedback rule from section~\ref{sec_coupling}.
As before, we prepare the system in its equilibrium state and switch on light spot~$1$ at time~$t=0$ which may lead to some symmetry breaking in the observed dynamic patterns.

In the following we discuss the observed dynamics of the light spots using the dynamic fingerprints introduced in Fig.~\ref{two_laser_control2}(c).
Neighboring rows correspond to neighboring spots in the polygon.
For all polygons animations of the switching light spots are included as movies M03 to M10 in the Supplemental Material.

\begin{figure}[tp]
\centering
\includegraphics[width=7.2cm]{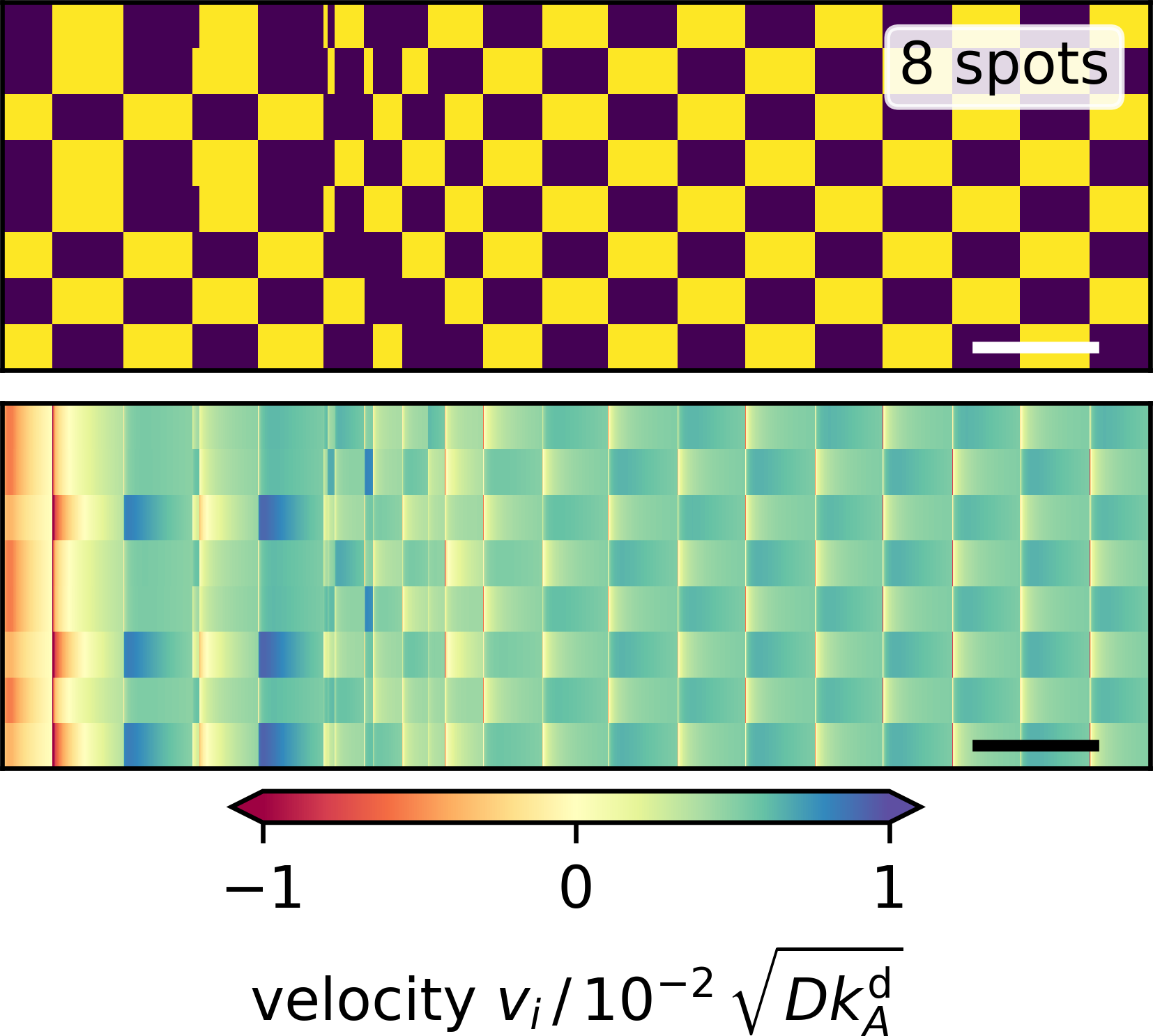}
\caption{
Dynamic fingerprints for an octagon of light spots.
Top:~'on' and 'off' states of the light spots versus time indicated by yellow and blue, respectively.
Each row represents one light spot numbered 1 to 8 from the bottom to the top.
Bottom:~radial velocity at each light spot plotted versus time with positive and negative values indicating outward and inward direction, respectively.
White and black bars in the bottom right corners represent the characteristic desorption time, i.e.~$1/k_A^\mathrm{d}$.
}
\label{plot_octagon}
\end{figure}

We first concentrate on the polygons with an even number of edges.
There is a strong anticorrelation between the on/off states of neighboring spots reminiscent of the oscillating state of two light spots.
So after some possible transient regime two sets of non-neighboring spots alternate with each other.
They create a checkerboard pattern in the dynamic fingerprint of the light spots as illustrated for $N=8$ in Fig.~\ref{plot_octagon}.
Polygons with varying $N$ differ in how the regular pattern is reached in time.
The square has a transient phase where the radial velocities at two opposing spots ($1$ and $3$) vary much more than at the other two spots and the dynamics does not follow a clear pattern before settling in two pairs of alternating spots.
The hexagon immediatly enters such a regular pattern with equal variation in all velocities.
The octagon (see Fig.~\ref{plot_octagon}) exhibits two transient phases:
At first, a triple and a quintuple of spots alternate followed by an irregular transition phase into the stable oscillatory pattern of two alternating quadruples.
In the triple-quintuple phase the velocities vary more at triple spots than at quintuple spots, which seems intuitive.

\begin{landscape}
\begin{figure}[tp]
\centering
\includegraphics[width=18cm]{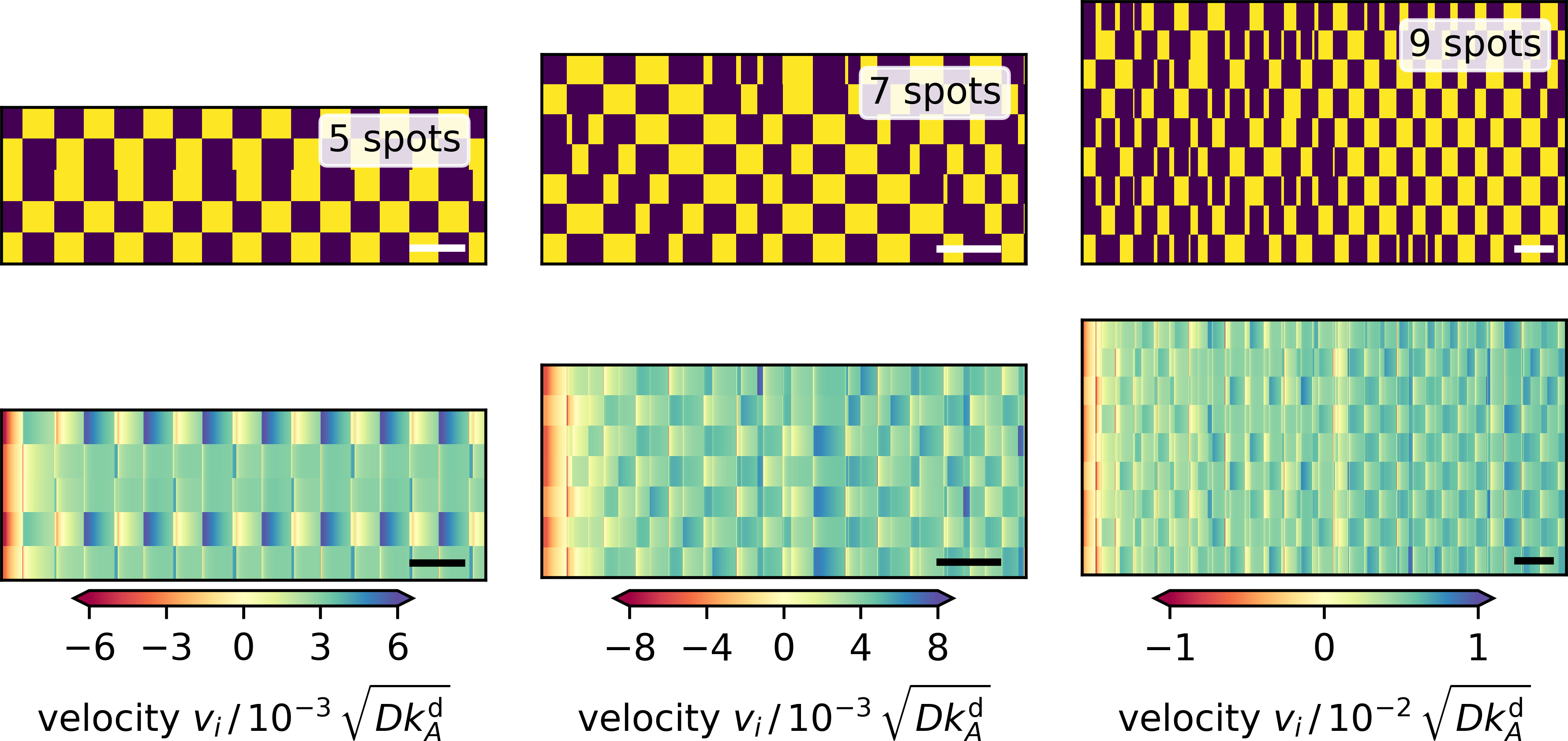}
\caption{
Dynamic fingerprints for 5, 7, and 9-sided polygons of light spots.
Top:~'on' and 'off' states of the light spots versus time indicated by yellow and blue, respectively.
Each row represents one light spot numbered 1 to $N$ from the bottom to the top.
Bottom:~radial velocity at each light spot plotted versus time with positive and negative values indicating outward and inward
direction, respectively.
White and black bars in the bottom right corners represent the characteristic desorption time~$1/k_A^\mathrm{d}$.
}
\label{plot_feedback_odd}
\end{figure}
\end{landscape}

In polygons with an odd number of edges, always two neighboring light spots are frustated since they cannot oscillate relative to each other.
So, we expect a more interesting dynamics.
The equilateral triangle of light spots (not shown) immediately settles into an oscillation pattern, where a single spot (spot~$1$) and the pair of the other spots alternate, since by symmetry the latter switch simultaneously.
The single spot is determined by the initial condition and there is a larger variation in radial velocity at spot~$1$, whereas the velocities at spots~$2$ and $3$ are similar.
The pentagon (Fig.~\ref{plot_feedback_odd}, left) behaves like the triangle:
It immediately reaches an oscillatory pattern, in which a pair and a triple of spots alternate.
Here, the radial velocity at the pair spots varies more strongly than at the triple spots.
The heptagon (Fig.~\ref{plot_feedback_odd}, center) is the first polygon with a long-lived irregular switching pattern,
which does not reach a regular oscillation state.
Notably, one observes some shorter on or off periods of spots, which propagate along neighbor-neighbor connections.
Finally, the nonagon (Fig.~\ref{plot_feedback_odd}, right) shows an irregular pattern as well and behaves similar to the heptagon, however the variations in the velocities are not as pronounced.
Thus, our investigations show that higher odd-numbered polygons settle into irregular patterns as suggested by the necessary frustration between neighboring spots.

\begin{figure}[tp]
\centering
\includegraphics[width=7cm]{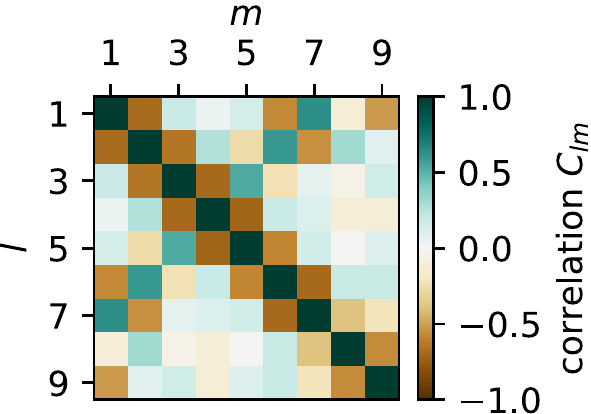}
\caption{
Color-coded correlation matrix~$C_{lm}$ of isomerization functions for all pairs of light spots~$l$ and $m$ for the on-off pattern of the nonagon in Fig.~\ref{plot_feedback_odd}.
Green indicates in-phase correlation, brown indicates anti-phase correlation, and dark shading (in both cases) indicates strong correlations.
}
\label{plot_corr09}
\end{figure}

To quantify the irregularity in the pattern of the nonagon, we calculate the correlation matrix~$C_{lm}$ between all pairs of light spots.
We define $C_{lm}$ for each respective pair of isomerization functions~$K_l$ and $K_m$:
\begin{equation}
C_{lm}
= \frac{\langle K_l, K_m \rangle}{\sqrt{\langle K_l, K_l \rangle
                                  \cdot \langle K_m, K_m \rangle}}
\end{equation}
with
\begin{equation}
\langle K_l, K_m \rangle
= \int \left(K_l(t) - \frac{K_{A\to B}}{2}\right)\left(K_m(t) - \frac{K_{A\to B}}{2}\right)\dif t
\end{equation}
The resulting pair-correlation matrix is displayed in Fig.~\ref{plot_corr09}.
It confirms that, while distant spots are uncorrelated, neighboring spots are anticorrelated.
In addition, it reveals some correlations between spots $1$ and $2$ and $6$ and $7$ which we expect to vanish in
the long-time limit.
The correlations between direct neighbors suggest a strong coupling between them, which is present in all studied polygons despite the fact that the projection angle in Eq.~(\ref{eq_projection}) approaches $90^\circ$ for large $N$.

Finally, we note for all polygons we observe that our feedback mechanism leads to net advection out of the polygons, which is indicated by the mainly blue regions in Figs.~\ref{plot_octagon} and \ref{plot_feedback_odd}.
So, in their centers fluid is drawn from the neighboring fluid phases towards the interface and the polygons of light spots act as pumps.

\section{Conclusions}
\label{sec_conclusions}
We have developed a theory for photoresponsive surfactants at a fluid-fluid interface and investigated their response to varying patterns of light.
In particular, we applied feedback control to generate them.
Our theory is formulated in a set of equations for the coarse-grained densities of each surfactant isomer and includes diffusive and advective Marangoni currents at the interface as well as ad-/desorption and photoisomerization.

We applied our theory to surfactants with isomers of identical desorption rates.
Their densities adjust to a single light spot diffusively, which reverses the direction of the radial Marangoni flow in time.
Inspired by this observation, we introduced a feedback coupling between the local advection velocities and isomerization rates of two light spots and thereby generated an alternating switching of the spots.
Dynamic fingerprints illustrate their temporal evolution.
When we extended the feedback coupling to light spots located on the vertices of regular polygons, we observed different dynamics for even and odd numbers of spots.
Even-sided polygons, after some possible transient regime, settle into periodically oscillating patterns, where two sets of next-nearest neighbor spots alternate with each other reminiscent of the oscillation in the two-spot system.
Odd-sided polygons with $3$ or $5$ vertices show similar behavior, however, two neighboring spots determined by inital conditions are always frustrated and switch approximately in phase.
Increasing the number of edges to $7$ or $9$, this frustration produces irregular patterns up to the end of our observation window.
The on/off states of nearest neighbor light spots are strongly anti-correlated, while distant spots show vanishing correlations.

Our findings provide a further example how feedback control can initiate complex spatio-temporal patterns in soft matter systems~\cite{prohm_feedback_2014,zeitz_feedback_2015}.
We will continue in this direction and investigate, for example, patterns, where the light spots are driven by time-delayed feedback~\cite{pyragas_continuous_1992}, which is known to either stabilize or destabilize nonlinear dynamic systems and to initiate complex patterns in space and time~\cite{schoell_handbook_2008}.
In this respect, regular polygons are a narrow class of networks, which we aim to extend to more general geometries with given ``connectivities'' between the light spots.

Our work combines the idea of  light-driven digital microfluidics with the concept of feedback control.
Overall, the implemented feedback rule generates flow fields with either regular or irregular oscillations.
Thereby, a self-regulated fluid pump occurs that moves fluid out of the polygons.
Since both pumping modes are produced by specific light patterns, rather than fixed geometric constraints, one can switch between both modes almost instantaneously and without mechanical adjustments.
We have restricted ourselves here to point-like spots of light.
However, different light patterns produced by moving light beams, holographic technique, and light interference provide further fascinating avenues of investigation.
Thus, varying light patterns in combination with photoresponsive surfactants offers the possibility of very flexible microfluidic applications~\cite{baigl_photo_2012}.

\subsubsection*{Conflicts of Interest}
There are no conflicts to declare.

\subsubsection*{Acknowledgements}
We thank J.~Blaschke, F.~Brückelmann, T.~Lubensky, and O.~Velev for interesting and helpful discussions.
J.\,G.~acknowledges funding from the German Research Foundation (DFG) via International Research Training Group~1524 and the hospitality of the University of Pennsylvania, Philadelphia, where part of the research was conducted.
Support from the Collaborative Research Center~910 is also acknowledged.

\newpage
\appendix

\section{Surface tension}
\label{appendix_surface_tension}
Surface tension~$\sigma$ is the thermodynamic conjugate force to surface area.
Therefore,
\begin{equation}
 \sigma=\partdif{F}{A} \, ,
\end{equation}
where $F$ is, e.g., free energy.
It is a homogeneous function, thus $F = A f$, where the free energy density $f$ only depends on particle densities~$\rho_A = $ and $\rho_B$.
Therefore we can write,
\begin{equation}
\sigma
= \partdif{(Af)}{A}
= f + A\partdif{f}{A}
= f + A \cdot \left( \partdif{f}{\rho_A} \partdif{\rho_A}{A}
  + \partdif{f}{\rho_B} \partdif{\rho_B}{A} \right)
\end{equation}
The particle densities are the ratios of particle numbers~$N_A$ and $N_B$ to $A$,
\begin{align}
 \rho_A &= \frac{N_A}{A} & \rho_B &= \frac{N_B}{A} \, ,
\end{align}
and we obtain
\begin{equation}
\sigma
  = f + A \cdot \left[ \partdif{f}{\rho_A} \left(-\frac{N_A}{A^2}\right)
  + \partdif{f}{\rho_B} \left(-\frac{N_B}{A^2}\right) \right]
\end{equation}
or after some simplifications:
\begin{equation}
\sigma = f - \rho_A \frac{\partial f}{\partial \rho_{A}} - \rho_B \frac{\partial f}{\partial \rho_{B}}
\end{equation}

\section{Connection to Flory-Huggins theory}
\label{flory_huggins}

A modern derivation of the Flory-Huggins free energy of a binary mixture can be found in Ref.~\autocite{hiemenz_polymer_2007} (Chapter 4).
We start with an analogous expression for ternary mixtures, where the third component is free space with local proportion~$\phi_\text{free}$,
\begin{multline}
\rlim^{-1} f
= \kbt \left( \phi_A \ln \phi_A + \phi_B \ln \phi_B + \phi_\mathrm{free} \ln \phi_\mathrm{free} \right)\\
+ \chi_{AB} \phi_A \phi_B + \chi_A \phi_A \phi_\mathrm{free} + \chi_B \phi_B \phi_\mathrm{free} \, .
\end{multline}
The entropic contribution from $\phi_\mathrm{free}=1-\phi_A-\phi_B$ effectively describes a hard-core interaction between components~$A$ and $B$:~its derivatives with respect to $\phi_{A/B}$ (the chemical potentials~$\mu_{A/B}$) diverge as $\phi_A + \phi_B$ approach 1.
Like the other interaction terms we approximate it to second order in $\phi_A$ and $\phi_B$ using $(1 - \phi) \ln (1 - \phi) \approx -\phi + 0.5\phi^2$ with $\phi = \phi_A + \phi_B$.
Regrouping terms, we find
\begin{multline}
\rlim^{-1} f
= \kbt \left[ \phi_A (\ln \phi_A - 1) + \phi_B (\ln \phi_B - 1) \right]
+ (\chi_{AB} - \chi_A - \chi_B + \kbt) \phi_A \phi_B
\\
- (\chi_A - \frac{1}{2}\kbt) \phi_A^2
- (\chi_B - \frac{1}{2}\kbt) \phi_B^2
+ \chi_A \phi_A + \chi_B \phi_B \, .
\end{multline}
In this form, $f$ is equivalent to Eq.~(\ref{eq_energy}), because $0$th and $1$st order terms do not contribute to
the gradients of surface tension or chemical potentials.
We can identify related quantities from both equations:
\begin{align}
\rho_A &= \rlim \phi_A &
\rho_B &= \rlim \phi_B
\nonumber \\
E_A &= \kbt - 2\chi_A &
E_B &= \kbt - 2\chi_B
\nonumber \\
E_{AB} &= (\chi_{AB} - \chi_A - \chi_B + \kbt)
\end{align}

\section{Oseen tensor integration by parts}
\label{app_oseen}
Consider Gauss's theorem applied to the following boun\-da\-ry integral on a compact region $\mathcal{A}\subset\mathbb{R}^2$ with boundary $\partial\mathcal{A}$ and boundary normal vector $\vect{n}\ofx$:
\begin{equation}
\oint\limits_{\partial\mathcal{A}} \sigma(\vect{x'}) \matr{O}(\vect{r-x'}) \vect{n}(\vect{x'}) \dif\vect{x'}
=
\int\limits_{\mathcal{A}} \nabla_{\vect{x'}}\cdot [ \matr{O}(\vect{r-x'})\sigma(\vect{x'})]\dif^2\vect{x'} \, .
\end{equation}
We apply the product rule to the right-hand side:
\begin{multline}
\oint\limits_{\partial\mathcal{A}} \sigma(\vect{x'}) \matr{O}(\vect{r-x'}) \vect{n}(\vect{x'}) \dif\vect{x'}
=
\int\limits_{\mathcal{A}} \sigma(\vect{x'})\nabla_{\vect{x'}} \matr{O}(\vect{r-x'}) \dif^2\vect{x'}
\\
  + \int\limits_{\mathcal{A}} \matr{O}(\vect{r-x'}) \nabla_{\vect{x'}}\sigma(\vect{x'}) \dif^2\vect{x'} \, .
\label{eq_gauss2}
\end{multline}
To show the equivalence of the second and third term, which correspond, respectively, to $\vect{v}$ from Eqs.~(\ref{eq_vectorform}) and (\ref{eq_oseen}), if $\mathcal{A}\to\mathbb{R}^2$, we need to prove that the boundary integral in Eq.~(\ref{eq_gauss2}) vanishes.
Consider the integral at a finite, but large distance $R$ from the center, where $\sigma\ofx=\sigma^\eq$:
\begin{equation}
\lim\limits_{R\to\infty} \oint\limits_{|\vect{x'}| = R}
  \sigma(\vect{x'}) \matr{O}(\vect{r - x'})
\frac{\vect{x'}}{|\vect{x'}|}
  \dif\vect{x'}
=
\sigma_\eq \lim\limits_{R\to\infty}
\oint\limits_{|\vect{x'}| = R}
  \matr{O}(\vect{r - x'})
  \frac{\vect{x'}}{|\vect{x'}|}
\dif\vect{x'} \, .
\end{equation}
Substituting $\matr{O}$ results in the integrand:
\begin{equation}
  \matr{O}(\vect{r - x'})
\frac{\vect{x'}}{|\vect{x'}|}
=
  \frac{\vect{r}\cdot\vect{x'} - |\vect{x'}|^2}{|\vect{x'}||\vect{r-x'}|^3} \vect{r}
  - \frac{\vect{r}\cdot\vect{x'} - |\vect{x'}|^2}{|\vect{x'}||\vect{r - x'}|^3} \vect{x'} \, .
\end{equation}
Most of the resulting integrals vanish as $R\to\infty$ because their integrands have order $|\vect{x'}|^{-2}$ or $|\vect{x'}|^{-3}$.
Only one integral is nontrivial because its integrand's order is $|\vect{x'}|^{-1}$.
We prove its convergence to zero by substituting $\vect{x'}=(R\cos\vartheta, R\sin\vartheta)$ explicitely:
\begin{equation}
\oint\limits_{|\vect{x'}| = R}
\frac{\vect{x'}|\vect{x'}|}{|\vect{r - x'}|^3}
\dif^2\vect{x'}
=
\int\limits_0^{2\pi}
\frac{
  (\cos\vartheta, \sin\vartheta)
}{
  (1 + (|\vect{r}|/R)^2 - 2\cos(\vartheta) x / R - 2\sin(\vartheta) y / R)^{3/2}
}
\dif\vartheta
\end{equation}
For any $\vect{r}=(x,y,z)$ the denominator of the integrand goes to $1$ as $R$ approaches infinity and, therefore,
\begin{equation}
\lim\limits_{R\to\infty}
\oint\limits_{|\vect{x'}| = R}
\frac{\vect{x'}|\vect{x'}|}{|\vect{r - x'}|^3}
\dif^2\vect{x'}
=
\int\limits_0^{2\pi}
\begin{pmatrix}
\cos(\vartheta)\\
\sin(\vartheta)
\end{pmatrix}
\dif\vartheta
= 0 \, .
\end{equation}
Thus, the boundary integral in Eq.~(\ref{eq_gauss2}) indeed vanishes.

\section{Estimating energy constants}
\label{parameter_extraction}

In their study of AzoTAB surfactants at an air-water interface Chevallier \textit{et al.}~provide surface tension measurements for two different $cis$-$trans$ mixing ratios at various bulk concentrations (Fig.~2 in~\autocite{chevallier_pumping_2011}).
They measured surface tension up to $100$ seconds after adding surfactants into their apparatus.
For large bulk concentrations their data show a clear saturation of surface tension in time.
We rely on these measured values of surface tension, which we show in the last column of Table~\ref{table_chevallier_data} together with the bulk concentrations of both surfactants (first and second column).
In addition, we use $\sigma_0=\SI{72.8}{\milli\newton / \meter}$ as the surface tension of a clean air-water interface at room temperature.

Chevallier \textit{et al.}~also provide a set of equations to map bulk concentrations $C^\eq_i$ to equilibrium values of surface densities $\rho^\eq_i$,
\begin{equation}
k^\mathrm{a}_i C^\eq_i (1 - \rho^\eq_i / \rlim) - k^\mathrm{d}_i \frac{\rho^\eq_i}{\rlim} = 0 \, .
\end{equation}
The surface densities calculated from these equations are displayed in Table~\ref{table_chevallier_data} (third and fourth column).
Finally, we determine the energy constants~$E_A$ and $E_{B}$ -- assuming $E_{AB}=(E_A + E_B)/2$ -- by inserting both sets of data points into Eq.~(\ref{eq_surface_tension}) for the surface tension.

\begin{table}
\centering
\begin{tabular*}{0.8\textwidth}{@{\extracolsep{\fill}}rrrrr}
\hline
$C^\eq_\mathrm{trans}$ &
$C^\eq_\mathrm{cis}$ &
$\rho^\eq_\mathrm{trans}$ &
$\rho^\eq_\mathrm{cis}$ &
$\sigma^\eq$
\\
\hline
$\SI{11.4}{\mol / \liter}$ &
$\SI{5.9}{\mol / \liter}$ &
$0.96\,\rlim$ &
$0.02\,\rlim$ &
$\SI{37}{\milli\newton / \meter}$
\\
$\SI{1.1}{\mol / \liter}$ &
$\SI{5.8}{\mol / \liter}$ &
$0.80\,\rlim$ &
$0.14\,\rlim$ &
$\SI{41}{\milli\newton / \meter}$
\\
\hline
\end{tabular*}
\caption{Our mapping from bulk concentrations~$C^\eq_i$ to surface densities~$\rho^\eq_i$ and surface tension measurements~$\sigma_\eq$ of data published by Chevallier et al.~in \autocite{chevallier_pumping_2011}.}
\label{table_chevallier_data}
\end{table}

\section{Rescaling length and time}
\label{app_char_quant}

In section \ref{sec_reduced} we reduce the dimensions of all physical quantities, using a set of characteristic parameters.
Some choices, like the use of $\rho^\infty$ for density or $\kbt$ for energy, follow naturally from the definition of $f$ in Eq.~(\ref{eq_energy}).
However, indentifying a characteristic length and time is not as straightforward.

Therefore, we consider an idealised system, where we assume density~$\rho_B$ to be sufficiently large so that during photoisomerization its value remains approximately constant.
Assuming also vanishing interactions between the surfactants ($E_A=E_B=E_{AB} = 0$), the free energy density~$f$ is
\begin{equation}
f = \sigma_0 + \kbt\rho_A [\ln(\lambda^2 \rho_A) - 1] \, ,
\end{equation}
where the constant contribution from the B surfactant is subsumed in $\sigma_0$.
Using Eq.~(\ref{eq_tension_general}) for surface tension~$\sigma$ and chemical potential $\mu_A = \partial_{\rho_A}f$, we find
\begin{equation}
\mu_A = \kbt\ln(\lambda^2 \rho_A)
\quad\text{and}\quad
\sigma = \sigma_0-\kbt\rho_A \, .
\end{equation}
We further introduce isomerization rates
\begin{equation}
k_{B\to A} = K_\mathrm{iso} \rho_B\delta\ofx
\quad\text{and}\quad
k_{A\to B} = 0 \, .
\end{equation}
Following the same procedure as in the main text, we arrive at a dif\-fu\-sion-ad\-vec\-tion-re\-ac\-tion equation for $\rho_A$,
\begin{equation}
\partdif{\rho_A}{t}
= D\Delta \rho_A
+ \frac{\kbt}{4\bar\eta}\nabla\cdot \left(\rho_A \nabla (-\Delta)^{-\frac{1}{2}}(\rho_A-\rho_A^\eq) \right)
- k^\mathrm{d}_A (\rho_A - \rho_A^\eq)
+ K_\mathrm{iso} \rho_B \delta\ofx \, .
\end{equation}
We linearize the equation by substituting $\rho_A\ofx=\rho_A^\eq + \delta\rho_A\ofx$ and by discarding all terms of second and higher order in $\delta\rho_A$:
\begin{equation}
\partdif{\delta\rho_A}{t}
= \left[
D\Delta
+ \frac{\kbt \rho_A^\eq}{4 \bar\eta} (-\Delta)^{\frac{1}{2}}
- k^\mathrm{d}_A
\right] \delta\rho_A\ofx
+ K_\mathrm{iso} \rho_B \delta\ofx \, .
\end{equation}
Taking the Fourier transform in both spatial dimensions, turns the partial differential equation into an ordinary differential equation:
\begin{equation}
\partdif{\delta\hat\rho_A}{t} = - \left(D |\vect{k}|^2 + \frac{\kbt \rho_A^\eq}{4\bar\eta}|\vect{k}|+ k^\mathrm{d}_A\right) \delta\hat\rho_A\ofk
+ K_\mathrm{iso} \rho_B \, .
\end{equation}
The steady-state solution is obtained by setting $\partial_t\delta\hat\rho_A=0$.
Note that $\delta\hat\rho_A$ is a naturally dimensionless quantity and we write:
\begin{equation}
\delta\hat\rho_A(\vect{k}, t\to\infty) = \frac{K_\mathrm{iso} \rho_B }{k^\mathrm{d}_A}\cdot \frac{1}{\frac{D}{k^\mathrm{d}_A} |\vect{k}|^2 + \frac{\kbt \rho_A^\eq}{4 k^\mathrm{d}_A \bar\eta}|\vect{k}|+ 1}
\end{equation}
We identify an amplitude $C=K_\mathrm{iso} \rho_B / k^\mathrm{d}_A$ and two lengths, the diffusion length~$l_\mathrm{D}=(D / k^\mathrm{d}_A)^{1/2}$ and the advection length~$l_\mathrm{A}=\kbt \rho_A^\eq (4 k^\mathrm{d}_A \bar\eta)^{-1}$.
In section \ref{sec_reduced} we combine $l_\mathrm{D}$ and $k^\mathrm{d}_A$ as characteristic quantities to eliminate~$D$ in the reduced equations.
The advection length~$l_\mathrm{A}$ is more situational because it depends on $\rho_A^\eq$.
Also note that $K_\mathrm{iso}$ only affects the amplitude of $\delta\hat\rho_A$ (and $\delta\rho_A$) but none of its length scales.

\section{Regridding scheme}
\label{app_regridding}

As noted in section~\ref{sec_solver}, FFTs are an efficient way to compute convolutions but they require orthogonal grids to operate on.
Therefore, we need to regrid data from the hexagonal grid onto an orthogonal grid to take advantage of FFT performance.

\begin{figure}
\centering
\includegraphics[width=9cm]{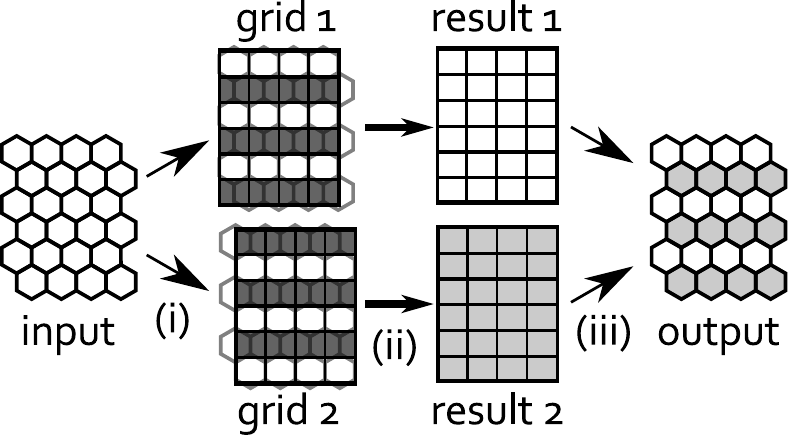}
\caption{Regridding procedure, from left to right: (i)~We divide input data into odd (grid 1) and even rows (grid 2).
The remaining rows (dark gray) in each grid are interpolated from input data.
(ii)~Convolutions are performed on both grids to produce results 1 and 2.
(iii)~Finally, we compile odd rows from result 1 with even rows from result 2 into output data.}
\label{schem_regrid}
\end{figure}

Figure~\ref{schem_regrid} shows two different orthogonal grids adjusted to the hexagonal lattice, where only in the white rows we can directly transfer the data from the center of the hexagons to the squares.
Thus we decided to use both grids for regridding our data.
To preserve as much accuracy as possible we interpolate the ``missing'' rows in each of those orthogonal grids using a higher-order polynomial interpolation method.
Note that we expect sharply peaked surface tension profiles because of the light spots and that, generally, there is a danger of oscillatory artifacts in interpolating peaked data.
We can avoid these artifacts by proceeding with \emph{both} grids and using the extra data to increase robustness.
With the interpolated rows we have constructed two complete orthogonal grids, on which we perform the FFT-based convolution.
After the convolution we recompile a hexagonal grid from both results taking only rows, which were not interpolated in the first regridding.
Because the response function in Eq.~(\ref{eq_mainresult}) is also peaked, this step corrects the previous interpolation and prevents oscillatory artifacts.

\newpage

\printbibliography[heading=bibintoc]

\end{document}